\documentclass[12pt]{iopart}

\usepackage{bm}
\usepackage{graphicx}
\usepackage{amsfonts}
\usepackage{amssymb}
\usepackage{amsbsy}
\usepackage{array}
\usepackage{iopams}
\usepackage{dsfont}
\usepackage{cite}

\begin{document}

\title{Phonon-induced dynamic resonance energy transfer}

\author{James Lim$^{1,2,3,4}$, Mark Tame$^{5,6}$, Ki Hyuk Yee$^{1,2}$, \\ Joong-Sung Lee$^{1,2}$ and Jinhyoung Lee$^{1,2}$}

\address{$^1$ Department of Physics, Hanyang University, Seoul 133-791, Korea}
\address{$^2$ Center for Macroscopic Quantum Control, Seoul National University, Seoul 151-742, Korea}
\address{$^3$ Research Institute for Natural Sciences, Hanyang University, Seoul 133-791, Korea}
\address{$^4$ Institute for Theoretical Physics, Albert-Einstein-Allee 11, University Ulm, D-89069 Ulm, Germany}
\address{$^5$ School of Chemistry and Physics, University of KwaZulu-Natal, Durban 4001, South Africa}
\address{$^6$ National Institute for Theoretical Physics, University of KwaZulu-Natal, Durban 4001, South Africa}

\ead{james.lim@uni-ulm.de, markstame@gmail.com and hyoung@hanyang.ac.kr}

\begin{abstract}

In a network of interacting quantum systems achieving fast coherent energy transfer is a challenging task. While quantum systems are susceptible to a wide range of environmental factors, in many physical settings their interactions with quantized vibrations, or phonons, of a supporting structure are the most prevalent. This leads to noise and decoherence in the network, ultimately impacting the energy-transfer process. In this work, we introduce a novel type of coherent energy-transfer mechanism for quantum systems, where phonon interactions are able to actually enhance the energy transfer. Here, a shared phonon interacts with the systems and dynamically adjusts their resonances, providing remarkable directionality combined with quantum speed-up. We call this mechanism phonon-induced dynamic resonance energy transfer and show that it enables long-range coherent energy transport even in highly disordered systems.

\end{abstract}


\maketitle

\newcommand{\bra}[1]{\left\langle #1\right|}
\newcommand{\ket}[1]{\left|#1\right\rangle}
\newcommand{\abs}[1]{\left|#1\right|}
\newcommand{\ave}[1]{\left<#1\right>}
\newcommand{\openone}{\leavevmode\hbox{\small1\normalsize\kern-.33em1}}


\section{Introduction}

Quantum systems exhibit many features that are contradictory to classical systems. The most well known are the concepts of complementarity, entanglement and nonlocality~\cite{Zeilinger1999}. Such nonclassical behavior enables quantum systems to outperform classical systems for a wide range of tasks in quantum information processing and communication~\cite{Nielsen2000}. In particular, in energy transport, quantum coherence enables a system to exploit the dynamics of quantum walks~\cite{Aharonov1993}, which are significantly faster than those of classical walks and diffusion. However, despite the many advantages offered by quantum systems, they are extremely challenging to isolate and invariably interact with their environment. This causes the loss of their quantum properties by decoherence, ultimately destroying any quantum advantage. In this work, we show that when quantum systems in a network are designed to interact with their environment in a shared way~\cite{Braun2002,Kim2002}, they are able to maintain their quantum coherence and even use it for achieving fast coherent energy transfer via quantum walk dynamics that are highly robust to disorder.

We consider quantized vibrational modes, or phonons, which are coupled to a network of two-level molecules where an electronic excitation of the molecules (exciton) is transferred. When electronic coherence rapidly decays by relaxation of a large number of phonon modes, or bath, the exciton tends to move downhill in energy through the molecules~\cite{Forster1948}. Here, the incoherent transfer of an exciton has previously been investigated with various phonon bath models, including local phonon baths, where each molecule is coupled to an independent phonon bath~\cite{Forster1948}, and a shared phonon bath, where different molecules are coupled to the same phonon modes~\cite{Soules1971}. However, recent advances in experimental techniques have made it possible to observe more complex behavior in the quantum regime for engineered structures~\cite{Collini2009,Eichenfield2009,OConnell2010,Semiao2010,Teufel2011,Chan2011,Hayes2013} and biological systems~\cite{Lee2007,Caram2012,Kolli2012,Chin2012,Chin2013,Susana2013}. For a shared phonon bath, the coherent transfer of an exciton has been investigated in the equilibrium regime, $\gamma\gg J$, where the relaxation rate $\gamma$ of phonons to an equilibrium state is faster than the dipole coupling $J$ between molecules~\cite{Nazir2009,McCutcheon2011}. Coherent exciton transfer has also been studied for the case that $\gamma\sim J$ where the phonons relax from a nonequilibrium to an equilibrium state during the exciton transfer~\cite{Wu2010,Ishizaki2010,HosseinNejad2010,Nalbach2010}. The nonequilibrium behavior of phonons has also been considered in the internal electron transfer in a molecule~\cite{Wolfseder1996}. In contrast to these earlier studies, here we investigate the exciton transfer through a chain of molecules in the nonequilibrium regime, $\gamma\ll J$, where the shared phonon is in a nonequilibrium state during the exciton transfer.

Open quantum systems have their internal energy levels shifted when they interact with an environment - the so-called Lamb shift, which takes place in a wide range of systems~\cite{Breuer2002}. The shift can even become time-dependent if the environment fluctuates temporally, {\it e.g.} the AC Stark effect, where oscillating electric fields cause a temporal shift in energy~\cite{Autler1955}. An example relevant to this work is that of a strongly coupled system of an exciton and a phonon. Here, the transfer of the exciton between dipole-interacting molecules causes the phonon to be in a nonequilibrium state, which leads to fluctuations in the energy levels of the molecules supporting the excitonic transfer. In this paper, we investigate a time-dependent Lamb shift induced by phonon modes that are shared by the interacting molecules, causing the energy-level fluctuations of the molecules to become highly correlated. We consider the non-adiabatic regime where the frequency of the phonons $\omega$ is comparable in magnitude to the dipole coupling $J$ between the molecules. Thus, we consider the nonequilibrium and non-adiabatic regime, $\gamma\ll J\sim \omega$, where the shared phonon is in a nonequilibrium state. We find that the resonance conditions for the exciton transfer are dynamically satisfied as a direct consequence of this time-dependent Lamb shift due to phonon dynamics. This leads to interesting energy transport features: (a) Excitonic hopping between molecules is dynamically conditioned (dynamic resonance), (b) The exciton transfer can be biased toward a particular direction (directionality), and (c) The exciton moves rapidly by interfering with itself (quantum walk). We show that this mechanism, which we call phonon-induced dynamic resonance energy transfer (DRET), enables long-range energy transport combined with quantum speed-up, even in the presence of considerable disorder in molecular energies. We also consider the case that $\gamma\sim J\sim \omega$ where DRET and incoherent transfer take place concurrently due to the relaxation of a shared phonon bath. We find that despite the detrimental effects of phonon relaxation and thermal noise, DRET can enhance exciton transfer in disordered systems by exploiting an interplay between quantum delocalization and classical funneling. Our work shows that sharing phonons between neighboring molecules not only protects quantum coherence~\cite{Lee2007,Nazir2009,McCutcheon2011,Wu2010,Ishizaki2010,HosseinNejad2010,Nalbach2010}, similar in spirit to decoherence-free subspaces~\cite{Lidar1998}, but also enables the use of quantum coherence for long-range energy transport in highly disordered systems. Our results highlight some interesting possibilities for future hybrid quantum phononic systems, with the features of the described mechanism potentially opening up new functionalities for designing quantum transport devices and networks.


\section{Exciton and shared phonon}\label{section_nonequilibrium_regime}

\begin{figure}
	\includegraphics{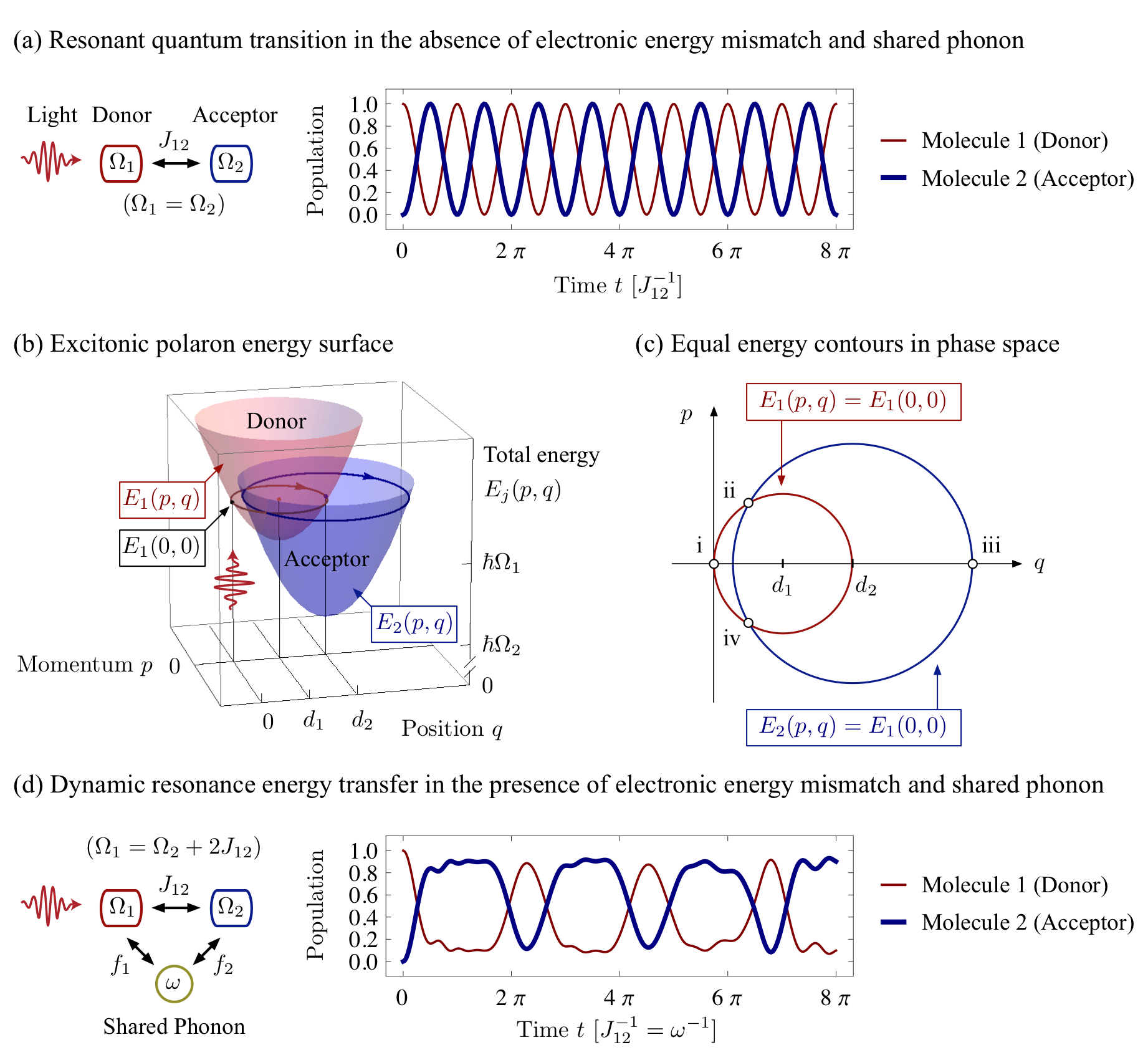}
	\caption{Dynamic resonance energy transfer in the presence of energy-level mismatch and shared phonon. We consider a two-molecule system with initial state $\ket{I}=\ket{1}$. In (a), where $(\Omega_1-\Omega_2)/J_{12}=0$ and $f_1=f_2=0$, resonant exciton hopping mediated by a dipolar coupling $J_{12}$ occurs and oscillates in time. In (b)-(d), where $\omega/J_{12}=1$, $(\Omega_1-\Omega_2)/J_{12}=2$, $f_1/J_{12}=1$ and $f_2/J_{12}=2$, a shared phonon in a nonequilibrium state leads to dynamic resonance energy transfer. A quantum picture of the phonon dynamics in phase space can be represented using the Wigner function~\cite{Wigner1932} (a video is included in the Supplemental Material).}
	\label{figure1}
\end{figure}

In both classical and quantum energy transport systems, resonance (energy-momentum matching) plays a vital role. In classical transport, with a phonon environment present, resonant incoherent energy transfer occurs for a downhill energy gradient, a mechanism well known as F{\"o}rster resonance energy transfer (FRET)~\cite{Forster1948}. Once an exciton hops incoherently from a high to a low energy molecule, the energy difference is dissipated by phonon relaxation, leading to funneling of the exciton through the molecules. However, in the presence of local energetic traps, due to disorder in the energy levels of the molecules, downhill diffusion leads to trapping of the exciton at the energetic minima, suppressing long-range energy transport. In quantum transport, on the other hand, resonant coherent energy transfer from a donor to an acceptor molecule occurs in the absence of energy-level mismatches, as shown in figure~\ref{figure1}(a). In the presence of disorder the transfer then becomes non-resonant and again long-range energy transport is suppressed.

We now show that when different molecules are coupled to the same phonon, the resonance conditions for quantum transport can be satisfied by the dynamics of the phonon, even in the presence of significant disorder in molecular energies. Such sharing of a phonon between molecules has been experimentally observed for engineered polymer systems~\cite{Collini2009} and natural photosynthetic complexes~\cite{Lee2007,Caram2012}. Consider the case where phonon relaxation is absent and the total energy of an excitonic polaron system is conserved. For a chain of $N$ molecules, the Hamiltonian of the system in the single-exciton manifold is
\begin{equation}
	\hat{H}=\sum_{j=1}^{N}\hat{E}_{j}(\hat{p},\hat{q})\ket{j}\bra{j}+\sum_{j\neq k}\hbar J_{jk}\ket{j}\bra{k},
\end{equation}
with $\ket{j}$ a single-exciton state where molecule $j$ is excited and all other molecules are in their ground states, $\hat{E}_{j}(\hat{p},\hat{q})$ is the excitonic polaron energy operator and $\hbar J_{jk}$ is the dipolar coupling responsible for exciton hopping. The operator $\hat{E}_{j}(\hat{p},\hat{q})$ is given by
\begin{equation}
	\hat{E}_{j}(\hat{p},\hat{q})=\hbar\Omega_j+\frac{\hat{p}^{2}}{2m}+\frac{1}{2}m\omega^2(\hat{q}-d_j)^{2},
\end{equation}
where $\hbar\Omega_j$ is the excited energy of molecule $j$ and $\omega$ is the frequency of the phonon mode associated with position $\hat{q}$ and momentum $\hat{p}$. The equilibrium position $d_j$ of the phonon mode is assumed to be deformed by the molecular excitation. Using the creation and annihilation operators of the phonon, $\hat{b}^{\dagger}$ and $\hat{b}$, the Hamiltonian can be decomposed as
\begin{equation}
	\hat{H}=\hat{H}_{e}+\hat{H}_{\rm ph}+\hat{H}_{e-\rm{ph}}.
\end{equation}
The first term describes the electronic states of the molecules
\begin{equation}
	\hat{H}_{e}=\sum_{j=1}^{N}\hbar(\Omega_{j}+f_{j}^{2}\omega^{-1})\ket{j}\bra{j}+\sum_{j\neq k}\hbar J_{jk}\ket{j}\bra{k},
\end{equation}
the second term describes the phonon
\begin{equation}
	\hat{H}_{\rm ph}=\hbar\omega(\hat{b}^{\dagger}\hat{b}+\frac{1}{2}),
\end{equation}
and the last term describes the interaction between exciton and phonon
\begin{equation}
	\hat{H}_{e-\rm{ph}}=-\sum_{j=1}^{N}\hbar f_j \ket{j}\bra{j}\otimes(\hat{b}^{\dagger}+\hat{b}),
\end{equation}
with $f_{j}=\sqrt{m\omega^{3}(2\hbar)^{-1}}d_{j}$, leading to fluctuations of the molecule energy levels. We assume that the total system is initially in the state $\ket{I}\otimes\ket{0}_{\rm ph}$ at time $t=0$, where $\ket{I}$ is the initial single-exciton state and $\ket{0}_{\rm ph}$ is the ground state of the phonon defined by $\hat{H}_{\rm ph}\ket{0}_{\rm ph}=\frac{1}{2}\hbar\omega\ket{0}_{\rm ph}$. The time evolution of the total system is $\sum_{k=1}^{N}\ket{k}\otimes\ket{\psi_{k}(t)}$, with $P_{k}(t)=\left<\psi_{k}(t)|\psi_{k}(t)\right>$ representing the population of molecule $k$ at time $t$.

\subsection{Dynamic resonance}

\begin{figure}
	\centering
	\includegraphics{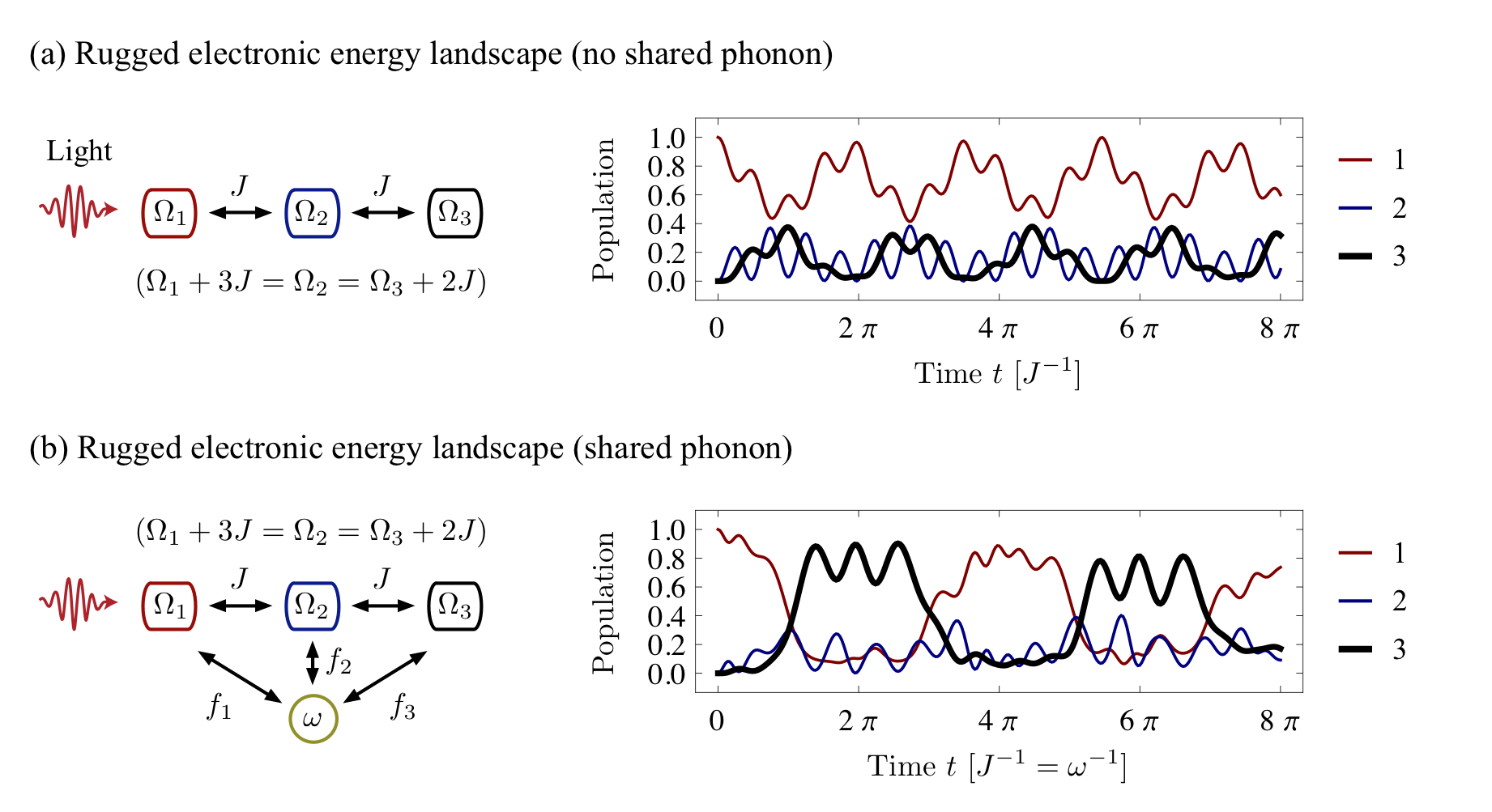}
	\caption{Energy transfer through a rugged uphill electronic energy landscape mediated by reversible exchange of energy between exciton and phonon. We consider a three-molecule system with initial state $\ket{I}=\ket{1}$ and take $J_{12}=J_{23}\equiv J$, $J_{13}=0$ (a linear chain), $(\Omega_{2}-\Omega_{1})/J=3$ and $(\Omega_{3}-\Omega_{1})/J=1$ (a rugged uphill electronic energy landscape, $\Omega_{2}>\Omega_{3}>\Omega_{1}$). In (a), where $f_{1}=f_{2}=f_{3}=0$, an exciton is localized at the initial molecule due to the non-resonance between molecules. In (b), where $\omega/J=1$, $f_{1}/J=2.11$, $f_{2}/J=2.80$ and $f_{3}/J=2.56$, a shared phonon leads to dynamic resonance energy transfer, such that an exciton can be efficiently transferred through the rugged landscape.}
	\label{figure2}
\end{figure}

To clarify the principle of DRET, we start with a two-molecule system. Here, once a donor is excited (molecule 1), the phonon mode has its equilibrium position displaced by $d_{1}$, due to a deformation in the structural support of the system. Then the phonon is in a nonequilibrium state, moving along an equal energy contour $E_1(p,q)=E_1(0,0)$ in phase space of momentum $p$ and position $q$, as shown in figures~\ref{figure1}(b) and (c). Here, $E_1(p,q)$ is the energy of the total system including the electronic energy of the donor $\hbar\Omega_1$ and the phonon energy. $E_1(0,0)$ is the initial energy of the total system. During the phonon's dynamics, if the equilibrium position of the phonon mode depends on the state of the molecules due to a different deformation in the structural support, {\it i.e.} $d_1\neq d_2$, $E_2(p,q)$ will oscillate in time as the difference between $E_1(p,q)$ and $E_2(p,q)$ is given by
\begin{equation}
\fl\quad\qquad	E_{1}(p,q)-E_{2}(p,q)=-m\omega^{2}(d_1-d_2)q+\hbar(\Omega_1-\Omega_2)+\frac{1}{2}m\omega^{2}(d_{1}^{2}-d_{2}^{2}),
\label{eq:Dynamic_L}
\end{equation}
which depends on the time evolution of $q$ when $d_{1}\neq d_{2}$. While the phonon moves along contour $E_1(p,q)=E_1(0,0)$, when $E_1(p,q)=E_2(p,q)$, the energy of the total system is conserved even if an exciton hops from the donor to the acceptor. This is the case at an intersection (point (ii) in figure 1(c)) between two contours $E_1(p,q)=E_1(0,0)$ and $E_2(p,q)=E_1(0,0)$. It is notable that the momentum $p$ is also conserved there. This energy-momentum matching is temporally conditioned by the phonon motion on two energy surfaces $E_1(p, q)$ and $E_2(p, q)$ ({\it cf}.~figures~\ref{figure1}(b) and (c)). We call such temporal matching of energy and momentum {\it dynamic resonance}. This is in contrast to static resonance when the donor and acceptor are always resonant (energy-matching) ({\it cf}.~figure~\ref{figure1}(a)). During the period of dynamic resonance there is a possibility to increase the hopping probability close to unity, depending on the dipolar coupling strength $J_{12}$, whereas otherwise such a possibility is prohibited. When the acceptor becomes excited, the phonon moves along contour $E_2(p,q)=E_1(0,0)$. This leads to temporal oscillations in $E_1(p,q)$ and non-resonance of the exciton transfer until the phonon arrives at another intersection (point (iv)). This phonon-induced dynamic resonance energy transfer (DRET) takes place even in the presence of electronic energy mismatches $\Omega_1\neq\Omega_2$. In the presence of the mismatch, the radii of equal energy contours are different, as shown in figure~\ref{figure1}(c), which leads to an asymmetry in the time evolution of the populations of the molecules, such that the exciton stays at the lower energy molecule 2 for longer, as shown in figure~\ref{figure1}(d).

It is remarkable that DRET is immune to variations of system parameters as far as the contours $E_1(p,q)=E_1(0,0)$ and $E_2(p,q)=E_1(0,0)$ intersect. For instance, in figure~\ref{figure1}(c), the red contour $E_1(p,q)=E_1(0,0)$ starts at the origin of the phase space when an exciton is initially created at molecule 1. The radius of the contour is thus determined by the equilibrium position $d_1$. The radius of the blue contour $E_2(p,q)=E_1(0,0)$ depends also on the electronic energy mismatch $\Delta\Omega_{12}=\Omega_1-\Omega_2$ so that it is larger than that of the red contour $E_1(p,q)=E_1(0,0)$ when $\Delta\Omega_{12}>0$ and vice versa. In figures~\ref{figure1}(b)--(d), where we take $\omega/J_{12}=1$, $f_{1}/J_{12}=1$ and $f_{2}/J_{12}=2$ ({\it cf.}~$f_{j}=\sqrt{m\omega^{3}(2\hbar)^{-1}}d_{j}$), the two contours remain intersecting over a wide range of electronic energy mismatches $-1\le\Delta\Omega_{12}/J_{12}\le 3$. If the mismatch is positive, {\it i.e.} $0<\Delta\Omega_{12}/J_{12}\le 3$, the radius of $E_2(p,q)=E_1(0,0)$ is larger than that of $E_1(p,q)=E_1(0,0)$, implying that the exciton stays longer at the lower energy molecule 2, which is the case of figure~\ref{figure1}(d) where we take $\Delta\Omega_{12}/J_{12}=2$.

In the presence of electronic energy mismatches, the energy of the excitonic polaron is partially transferred from the exciton to the phonon and back again without dissipation in order to allow the transfer to take place. This reversible exchange of energy enables efficient energy transfer even in rugged electronic energy landscapes with energetic traps, as shown in figure~\ref{figure2}. Here we consider a three-molecule linear chain with $J_{12}=J_{23}=J$ and $J_{13}=0$ where an exciton is initially localized at molecule 1 and it is transferred through a rugged electronic energy landscape with $\Omega_2>\Omega_3>\Omega_1$, implying that molecule 2 is an energetic barrier and molecules 1 and 3 are nonresonant. In the absence of a shared phonon, an exciton is localized at the initial molecule in a rugged landscape due to the non-resonance arising from the energy-level mismatches between initial molecule 1 and the other molecules, as shown in figure~\ref{figure2}(a). In the presence of a shared phonon, on the other hand, when the exciton hops from low energy molecule 1 to high energy molecule 2, the electronic energy is increased ($\Omega_1<\Omega_2$), while the potential energy of the phonon is decreased $\frac{1}{2}m\omega^2(q-d_1)^2>\frac{1}{2}m\omega^2(q-d_2)^2$ due to the change of the equilibrium position of the phonon mode ($d_1\rightarrow d_2$). In an opposite way, when the exciton hops from high energy molecule 2 to low energy molecule 3, the electronic energy is decreased ($\Omega_2>\Omega_3$), while the potential energy of the phonon is increased. With energy conservation satisfied, quantum coherence then enables a unidirectional exciton transfer from molecule 1 to 3, such that the exciton stays at molecule 3 for a while with a high probability, as shown in figure~\ref{figure2}(b).

\subsection{Directionality}

\begin{figure}
	\includegraphics{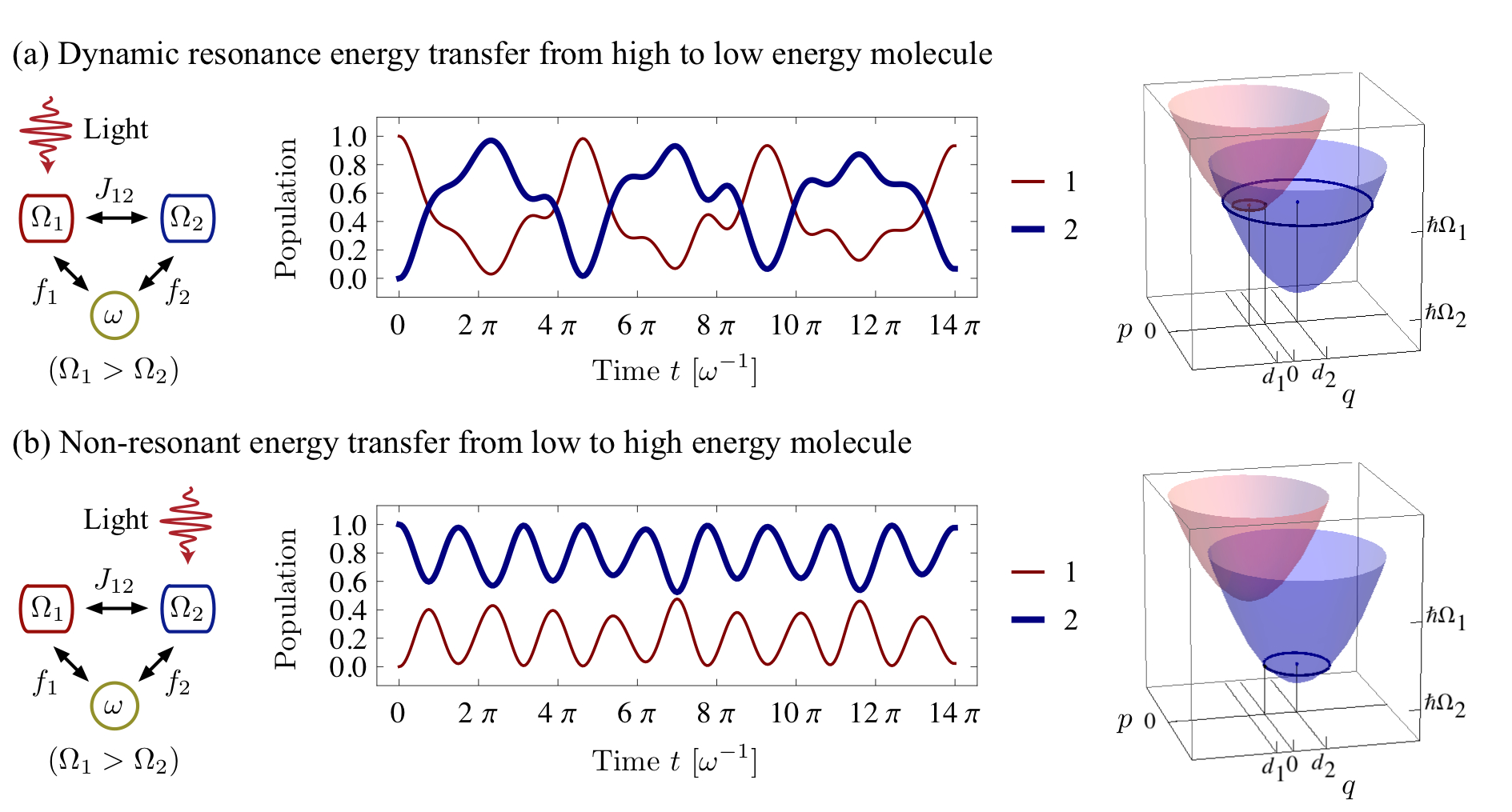}
	\caption{Directionality of downhill energy transfer induced by dynamic resonance. Here we consider a two-molecule system and take $\omega/J_{12}=2.4$, $(\Omega_{1}-\Omega_{2})/J_{12}=2$, $f_1/J_{12}=-0.5$ and $f_2/J_{12}=1$. In (a), where $\ket{I}=\ket{1}$, an exciton created at molecule 1 is transferred from a high to a low energy molecule by dynamic resonance. In (b), where $\ket{I}=\ket{2}$, for the same choice of system parameters, an exciton created at molecule 2 is localized at the low energy molecule 2 by non-resonance.}
	\label{figure3}
\end{figure}

\begin{figure}
	\includegraphics{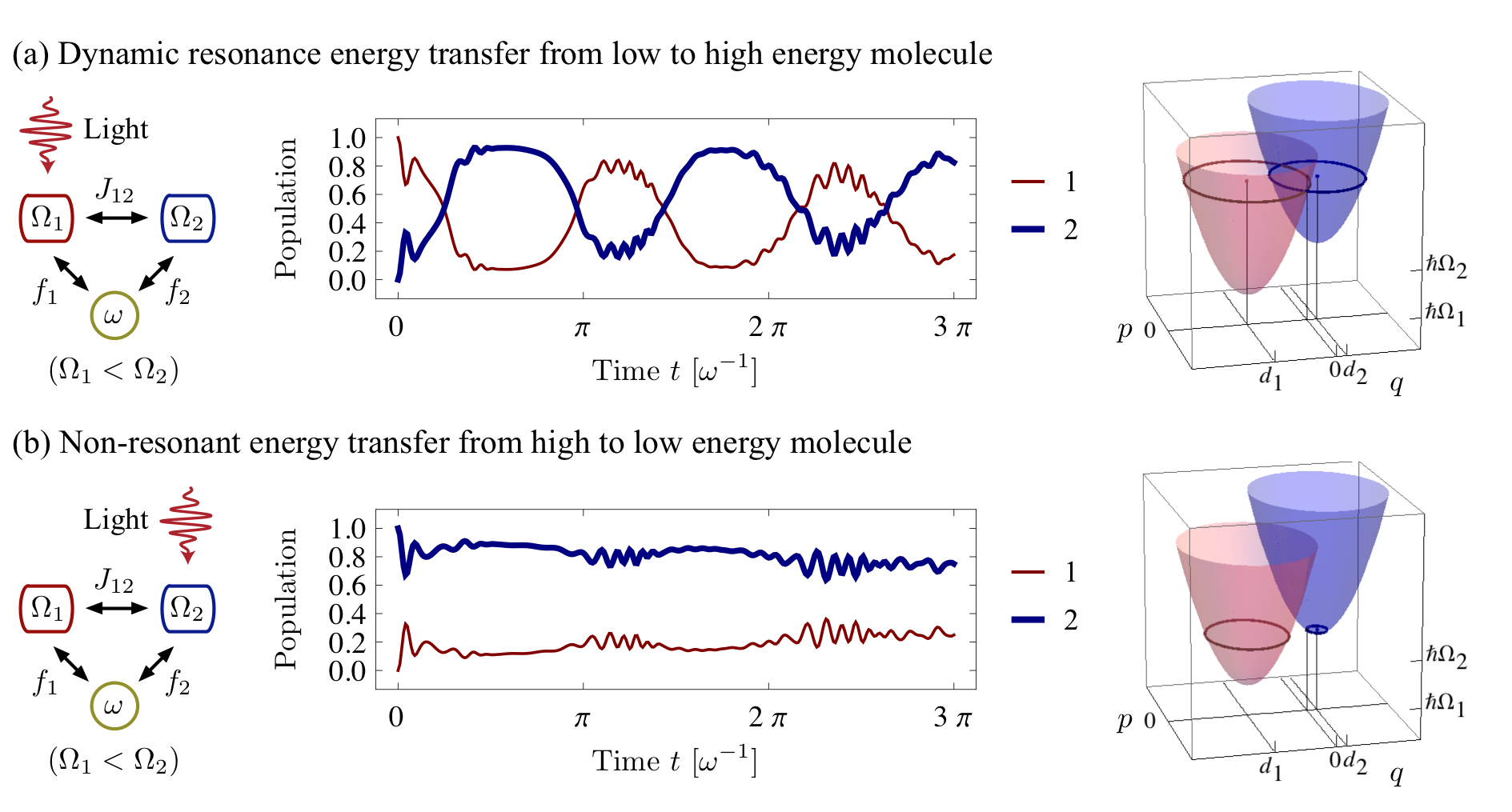}
	\caption{Directionality of uphill energy transfer induced by dynamic resonance. Here we consider a two-molecule system and take $\omega/J_{12}=0.143$, $(\Omega_{1}-\Omega_{2})/J_{12}=-2.14$, $f_{1}/J_{12}=-0.857$ and $f_{2}/J_{12}=0.143$. In (a), where $\ket{I}=\ket{1}$, an exciton created at molecule 1 is transferred from a low to a high energy molecule by dynamic resonance. In (b), where $\ket{I}=\ket{2}$, an exciton created at molecule 2 is localized at the high energy molecule 2 by non-resonance.}
	\label{figure4}
\end{figure}

\begin{figure}
\fl\qquad\qquad~\includegraphics{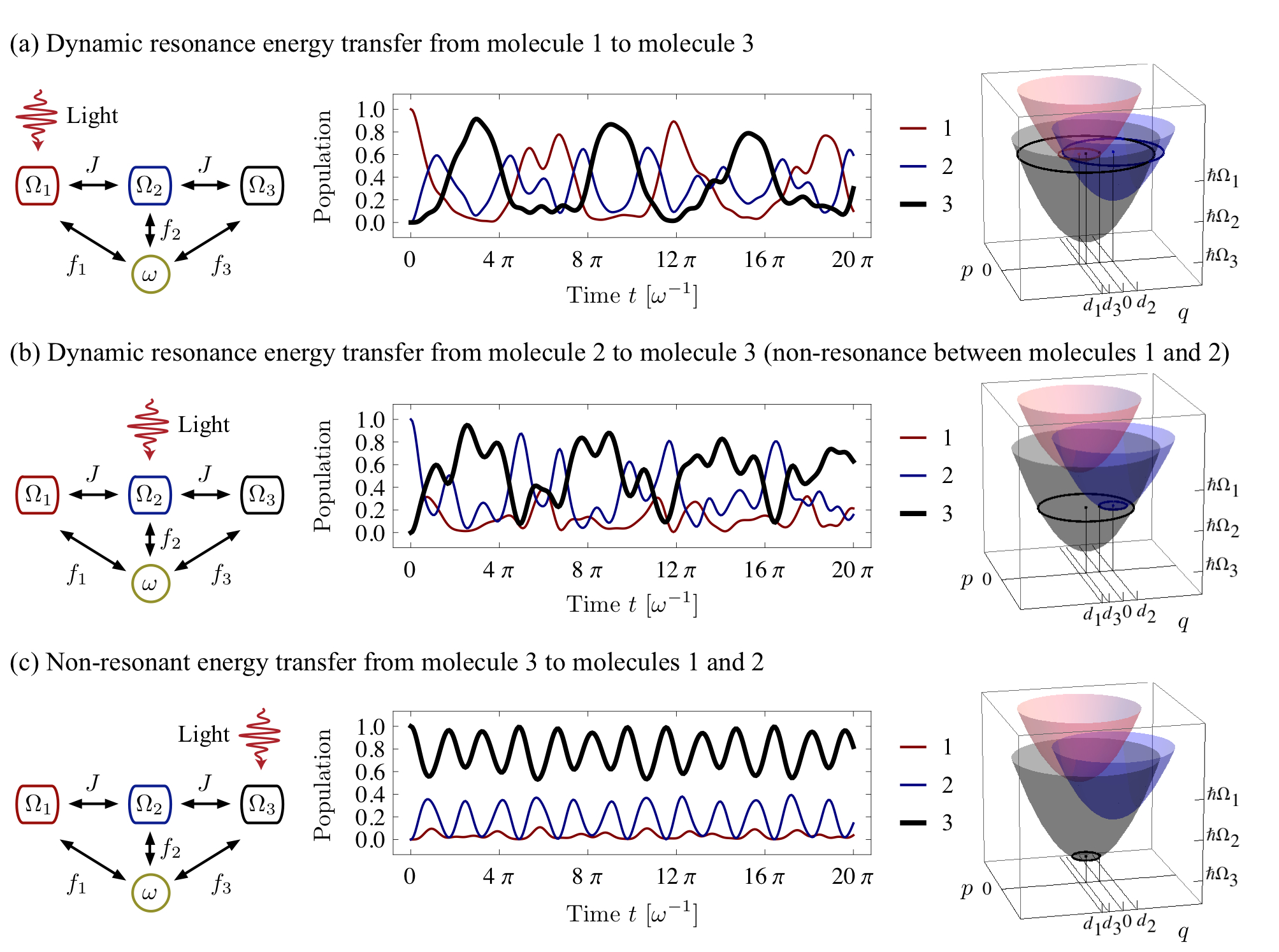}
	\caption{Directionality of dynamic resonance energy transfer in a chain of molecules. Here we consider a three-molecule linear chain and take $J_{12}=J_{23}\equiv J$, $J_{13}=0$ (a linear chain), $(\Omega_{1}-\Omega_{3})/J=4$, $(\Omega_{2}-\Omega_{3})/J=2$ (a downhill electronic energy landscape, $\Omega_{1}>\Omega_{2}>\Omega_{3}$), $\omega/J=2.29$, $f_{1}/J=-0.975$, $f_{2}/J=0.654$ and $f_{3}=-0.654$. In (a), (b) and (c), we take $\ket{I}=\ket{1}$, $\ket{I}=\ket{2}$ and $\ket{I}=\ket{3}$ respectively.}
	\label{figure5}
\end{figure}

Remarkably, DRET can also induce directionality, similar to a diode in electrical circuitry. Once an exciton is created at molecule $j$, the equal energy contours $E_{k}(p,q)=E_{j}(0,0)$ are determined in phase space. If the contours are close enough such that the total energy difference is small compared to the dipolar coupling, $\abs{E_{k}(p,q)-E_{l}(p,q)}<\hbar J_{kl}$, near-resonant exciton transfer takes place. If not, the exciton transfer is suppressed by non-resonance. This can induce directionality, such that energy transfer in a particular direction is enhanced by dynamic resonance, while in the opposite direction it is suppressed by non-resonance.

In direct contrast to FRET, where the energy transfer is always biased toward a lower energy molecule, DRET can induce directionality in both downhill and uphill directions. In figure~\ref{figure3}, we consider a two-molecule system with  $\Omega_1>\Omega_2$ where molecule 1 has higher electronic energy than molecule 2. Here we take system parameters such that the equal energy contours $E_{1}(p,q)=E_{1}(0,0)$ and $E_{2}(p,q)=E_{1}(0,0)$ are close enough when high energy molecule 1 is excited, while the contours $E_{1}(p,q)=E_{2}(0,0)$ and $E_{2}(p,q)=E_{2}(0,0)$ have no intersections when low energy molecule 2 is excited. This can be achieved by adjusting the equilibrium positions $d_j$ and frequency $\omega$ of the phonon mode, which determine the location and shape of energy surfaces $E_{1}(p,q)$ and $E_{2}(p,q)$ ({\it cf.}~figure~\ref{figure1}(b)). In this case, the exciton transfer from the high to the low energy molecule is enhanced by dynamic resonance, as shown in figure~\ref{figure3}(a), while the transfer from the low to the high energy molecule is suppressed by non-resonance, as shown in figure~\ref{figure3}(b).  On the other hand, in figure~\ref{figure4}, we consider a two-molecule system with $\Omega_1<\Omega_2$ where molecule 1 has lower electronic energy than molecule 2. Here we take another set of system parameters such that the equal energy contours $E_{1}(p,q)=E_{1}(0,0)$ and $E_{2}(p,q)=E_{1}(0,0)$ have intersections when low energy molecule 1 is excited, while the contours $E_{1}(p,q)=E_{2}(0,0)$ and $E_{2}(p,q)=E_{2}(0,0)$ are far enough apart when high energy molecule 2 is excited. In this case, contrary to figure~\ref{figure3} the exciton transfer from the low to the high energy molecule is enhanced by dynamic resonance, as shown in figure~\ref{figure4}(a), while the transfer from the high to the low energy molecule is suppressed by non-resonance, as shown in figure~\ref{figure4}(b).

Directionality can also be induced in a chain of molecules. In figure~\ref{figure5}, we consider a three-molecule linear chain with $J_{12}=J_{23}=J$ and $J_{13}=0$ where an exciton is transferred through a downhill electronic energy landscape with $\Omega_1>\Omega_2>\Omega_3$. Here we take system parameters such that downhill energy transfer is enhanced by dynamic resonance, while uphill transfer is suppressed by non-resonance. In figure~\ref{figure5}(a), an exciton created at molecule 1 is resonantly transferred to lower energy molecules by dynamic resonance. In figure~\ref{figure5}(b), on the other hand, an exciton created at molecule 2 is resonantly transferred to lower energy molecule 3 by dynamic resonance, while the exciton transfer to higher energy molecule 1 is suppressed by non-resonance. In figure~\ref{figure5}(c), an exciton created at molecule 3 is localized at  the initial molecule with a high probability due to the non-resonance between molecule 3 and the other molecules.

Recently directionality in quantum systems has been studied in the context of discrete-time open quantum walks~\cite{Attal2012,Sinayskiy2013}. It has been shown that a quantum walk can be driven by the interaction with an environment such that directionality takes place in quantum transport. This can be realized using various operations on quantum system and environment in each time step, such as swapping the state of the system and that of the environment and then re-preparing an environment in a particular state. This scenario is contrary to DRET, which does not require such operations on an environment (phonon) during system dynamics (exciton transfer).

\subsection{Quantum walk}

\begin{figure}
	\includegraphics{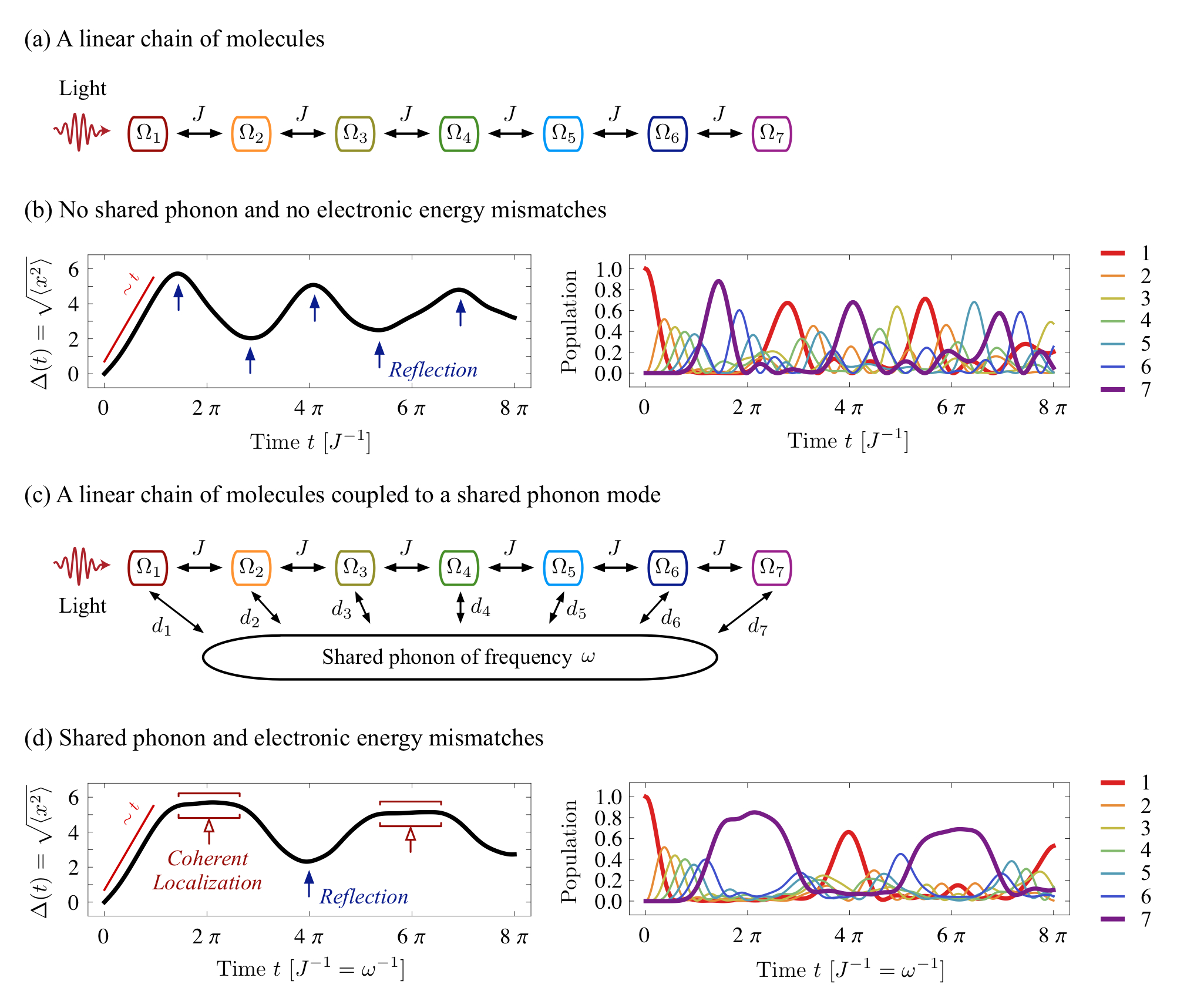}
	\caption{Quantum walk and coherent localization induced by a shared phonon mode. In (a), a linear chain of molecules is shown with uniform dipolar couplings $J$ between neighboring molecules ($J_{jk}\equiv J$ if $j=k\pm 1$ and $J_{jk}=0$ otherwise). Here we take $\ket{I}=\ket{1}$, $\Omega_{j}-\Omega_{k}=0$ for all $j$ and $k$ (a flat electronic energy landscape), and $f_{j}=0$ for all $j$ (no interaction between molecules and phonon mode). In (b), the dynamics of the populations of the molecules are shown, which leads to the root mean square displacement $\Delta(t)$. In (c), a linear chain of molecules is coupled to a shared phonon mode. Here we take $\ket{I}=\ket{1}$, $(\Omega_{1}-\Omega_{7})/J=1.93$, $(\Omega_{2}-\Omega_{7})/J=2.06$, $(\Omega_{3}-\Omega_{7})/J=2.11$, $(\Omega_{4}-\Omega_{7})/J=2.13$, $(\Omega_{5}-\Omega_{7})/J=2.14$, $(\Omega_{6}-\Omega_{7})/J=2.05$, $\omega/J=1$, $f_{1}/J=-0.471$, $f_{2}/J=-0.305$, $f_{3}/J=-0.221$, $f_{4}/J=-0.151$, $f_{5}/J=0.129$, $f_{6}/J=0.325$ and $f_{7}/J=1.47$. In (d), the dynamics of the populations of the molecules are shown, which leads to the root mean square displacement $\Delta(t)$.}
	\label{figure6}
\end{figure}

In a large chain, DRET also enables long-range energy transport that is otherwise unavailable. In the absence of energy mismatches, a coherent exciton moves very quickly by interfering with itself similar to a quantum walk. In figure~\ref{figure6}(a), a linear chain of molecules is shown with uniform dipolar couplings $J$ between neighboring molecules. In the absence of energy mismatches, a resonant quantum transition takes place between the molecules, as shown in figure~\ref{figure6}(b). This leads to a quantum walk in a chain of finite size where an exciton is transferred back and forth through the molecules with quantum speed-up due to the reflection at the boundaries of the chain. Here the root mean square displacement $\Delta(t)$ increases linearly with time $t$
\begin{equation}
	\Delta(t)=\sqrt{\ave{x^2}}=\left(\sum_{k=1}^{7}(k-1)^2 P_{k}(t)\right)^{1/2}\propto t,
\end{equation}
where $x$ is the distance that the exciton moves during time $t$~\cite{Aharonov1993}. This is quadratically faster than the root mean square displacement of a classical walk, $\Delta(t)\propto\sqrt{t}$. In this work, we say that {\it quantum delocalization} takes place in exciton transfer when its root mean square displacement increases linearly with time, similar to a quantum walk. In the presence of energy mismatches, however, quantum coherence leads to coherent localization of the exciton, similar to Anderson localization~\cite{Anderson1958}. A new and rather surprising feature introduced by DRET is that when electronic energy mismatches are present and a shared phonon is available, both quantum walk and coherent localization can occur concurrently, where an exciton moves through a chain with quantum speed-up and is subsequently localized at a particular molecule for a period without losing quantum coherence. In figure~\ref{figure6}(c), a linear chain of molecules is coupled to a shared phonon mode. Even in the presence of electronic energy mismatches, an exciton can be transferred with quantum speed-up by dynamic resonance, as shown in figure~\ref{figure6}(d). When molecule 7, located at one of the boundaries, has the lowest electronic energy, the exciton is temporally trapped at the molecule due to the non-resonance induced by the phonon dynamics.


\section{Dynamic resonance under phonon relaxation and thermal noise}\label{section_quasiequilibrium_regime}

So far only a single shared phonon mode has been considered. We now consider a molecular chain coupled to a shared phonon bath modeled by a set of quantum harmonic oscillators to investigate the influence of phonon relaxation and thermal noise on dynamic resonance energy transfer. The Hamiltonian in the single-exciton manifold consists of three parts $\hat{H}=\hat{H}_{e}+\hat{H}_{\rm ph}+\hat{H}_{e-{\rm ph}}$. The first term describes the electronic states of the molecules
\begin{equation}
	\hat{H}_{e}=\sum_{j=1}^{N}\hbar(\Omega_{j}+\lambda_{j}(1+2s_{j}))\ket{j}\bra{j}+\sum_{j\neq k}\hbar J_{jk}\ket{j}\bra{k},
	\label{eq:H_e}
\end{equation}
where $\lambda_{j}=\sum_{\xi}g^{2}_{j\xi}\omega_{\xi}^{-1}$, $\lambda_{j}(1+2s_{j})$ is the Lamb shift induced by the interaction between molecule $j$ and the phonons. The second term describes the phonons $\hat{H}_{\rm ph}=\sum_{\xi}\hbar\omega_{\xi}\hat{b}^{\dagger}_{\xi}\hat{b}_{\xi}$, where $\hat{b}^{\dagger}_{\xi}$ and $\hat{b}_{\xi}$ are creation and annihilation operators of a phonon mode $\xi$ with frequency $\omega_{\xi}$. The last term describes the interaction between the molecules and phonons
\begin{equation}
	\hat{H}_{e-{\rm ph}}=-\sum_{j=1}^{N}\sum_{\xi}\hbar g_{j\xi}(\ket{j}\bra{j}+s_{j}\sum_{k=1}^{N}\ket{k}\bra{k})\otimes(\hat{b}^{\dagger}_{\xi}+\hat{b}_{\xi}).
	\label{eq:H_e_ph}
\end{equation}
When $s_{j}=0$ for all $j$, $\hat{H}_{e-{\rm ph}}$ describes local phonon baths: for a phonon mode $\xi$ coupled to molecule $j$ by $g_{j\xi}\neq 0$, we assume $g_{k\xi}=0$ for all $k\neq j$. For nonzero $s_{j}$, $\hat{H}_{e-{\rm ph}}$ describes a shared phonon bath where each phonon mode $\xi$ is coupled to all the molecules by $g_{j\xi}s_{j}$, except for molecule $j$ which is coupled by $g_{j\xi}(1+s_{j})$. In this work, we take an Ohmic spectral density with a Lorentz-Drude cutoff function
\begin{equation}
	\Lambda_{j}(\omega)=\sum_{\xi}\hbar g_{j\xi}^{2}\delta(\omega-\omega_{\xi})=\frac{2\hbar\lambda_{j}}{\pi}\frac{\omega\gamma_{j}}{\omega^{2}+\gamma_{j}^{2}},
	\label{eq:LD}
\end{equation}
where $\gamma_{j}$ is the phonon relaxation rate.

The dynamics of the exciton transfer are described by the reduced electronic state $\hat{\rho}_{e}(t)={\rm Tr_{ph}}[\hat{\rho}(t)]$, where $\hat{\rho}(t)$ is the total state and ${\rm Tr_{ph}}$ is the partial trace over phonons. We take the state of the total system at time $t=0$ to be $\hat{\rho}(0)=\hat{\rho}_{e}(0)\otimes Z^{-1}\exp(-\beta\hat{H}_{\rm ph})$, where $\hat{\rho}_{e}(0)$ is the initial single-exciton state and $Z^{-1}\exp(-\beta\hat{H}_{\rm ph})$ is a thermal state at temperature $T=(k_{B}\beta)^{-1}$, with $Z={\rm Tr_{ph}}[\exp(-\beta\hat{H}_{\rm ph})]$.

In appendix~A, we show that the shared phonon bath induces dynamic resonance such that the effective energy landscape that an exciton feels during transport changes in time $\hbar\Omega_{j}+\hbar\lambda_{j}(1+2s_{j}e^{-\gamma_{j}t})$: the effective energy level of molecule $j$ is $\hbar\Omega_{j}+\hbar\lambda_{j}(1+2s_{j})$ at the initial time $t=0$ and it converges to $\hbar\Omega_{j}+\hbar\lambda_{j}$ as the time-dependent Lamb shift $2\hbar\lambda_{j}s_{j}e^{-\gamma_{j}t}$ decays with the phonon relaxation rate $\gamma_{j}$. It is notable that the time-dependent Lamb shift vanishes when $s_{j}=0$ for all $j$, which is the case of the local phonon baths. In appendix B, we provide a non-Markovian quantum master equation that reliably describes exciton transfer dynamics in the presence of the time-dependent Lamb shift.

\subsection{Quantum delocalization and classical funneling}

\begin{figure}
\fl\qquad\quad\includegraphics{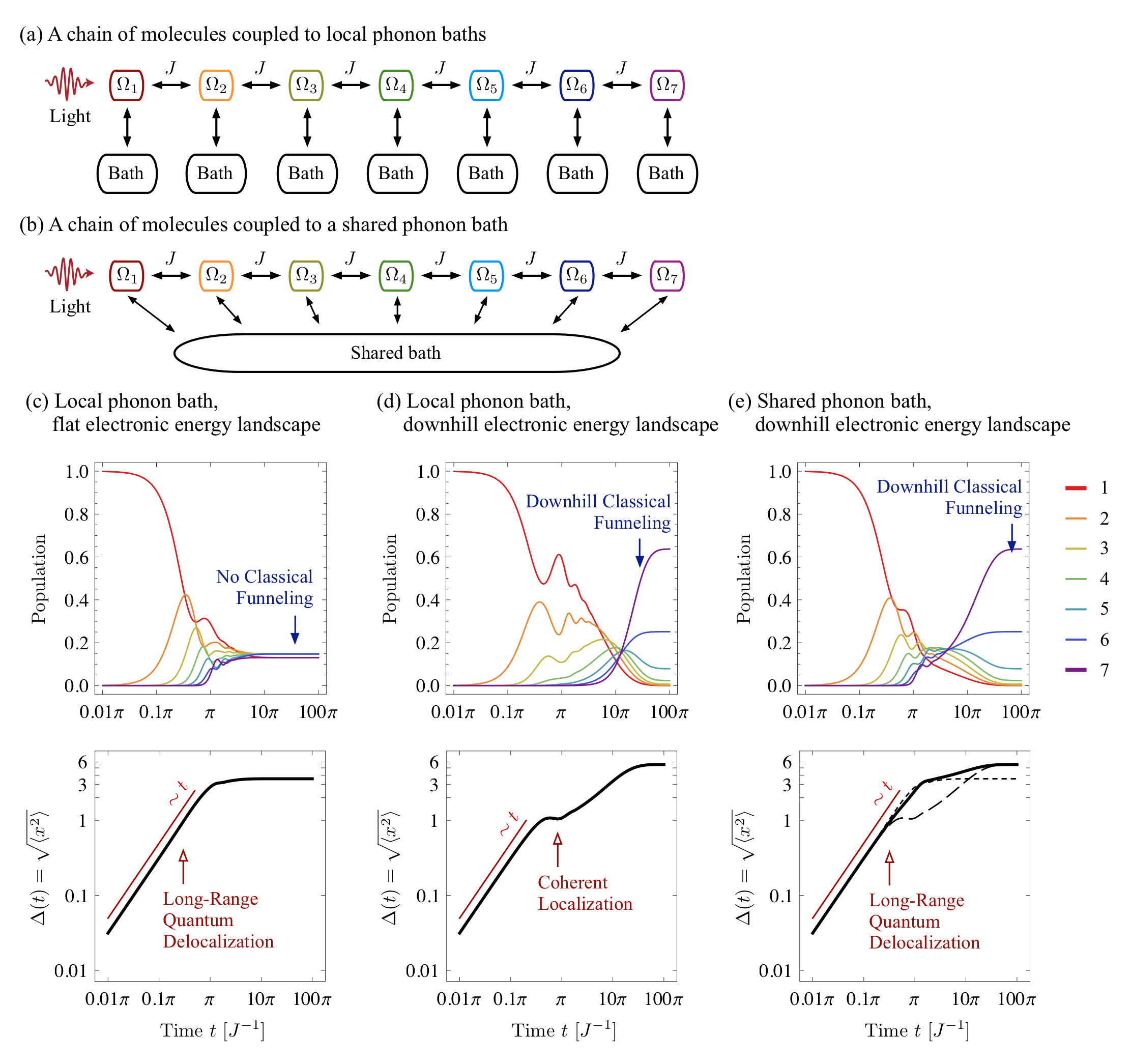}
	\caption{Long-range quantum delocalization and classical funneling induced by a shared phonon bath. In (a), a linear chain of molecules is shown, where each molecule is coupled to an independent phonon bath. In (b), a linear chain coupled to a shared phonon bath is shown. In (c)-(e), where $\hat{\rho}_{e}(0)=\ket{1}\bra{1}$, $J_{jk}\equiv J$ if $j=k\pm 1$ and $J_{jk}=0$ otherwise, $\lambda_{j}/J=0.35$, $\gamma_{j}/J=0.35$, $k_{B}T/\hbar J=2$, the dynamics of the populations of the molecules and the corresponding root mean square displacement $\Delta(t)$ are shown. In (c), we take $(\Omega_{j}-\Omega_{k})/J=0$ for all $j$ and $k$ (a flat electronic energy landscape)  and $s_{j}=0$ for all $j$ (local phonon baths). In (d), we take $(\Omega_{j}-\Omega_{j+1})/J=2$ (a downhill electronic energy landscape) and $s_{j}=0$ for all $j$ (local phonon baths). In (e), we take $(\Omega_{j}-\Omega_{j+1})/J=2$ (a downhill electronic energy landscape) and $s_{j}=j\times (k_{B}T/\hbar\lambda_{j})$ (a shared phonon bath).}
	\label{figure7}
\end{figure}

As the number of shared phonon modes is increased leading to a shared bath, phonon relaxation takes place, suppressing the reversibility of DRET. Here, once an exciton hops from a high to low energy molecule by dynamic resonance, it is trapped at the lower energy molecule due to phonon relaxation, enhancing the directionality of the exciton transfer. On the other hand, in the presence of thermal noise, DRET is degraded as quantum coherent effects are suppressed by decoherence and the trapped exciton at the lower energy molecule may be released by absorbing thermal energy. Despite the detrimental effects of phonon relaxation and thermal noise, DRET can enhance exciton transfer in disordered systems by exploiting an interplay between quantum delocalization and classical funneling. Using the effective Hamiltonian model developed in the appendix (based on equations~\eref{eq:H_e} and \eref{eq:H_e_ph}) we can calculate the density matrix $\hat{\rho}_{e}(t)$ for the reduced electronic system. In figure~\ref{figure7}, as an example, we consider a seven-molecule linear chain coupled to phonon baths. When an exciton is created at molecule 1, it is transferred through the molecules, while thermal noise of the phonon bath at temperature $T$ destroys quantum coherence, leading to a quantum-to-classical transition in the exciton transfer. Here we take environment couplings such that quantum transport takes place up to $t\approx\pi J^{-1}$ and classical transport occurs subsequently. In the presence of local baths, as shown in figure~\ref{figure7}(a), where no time-dependent Lamb shift takes place, when all the molecules have the same electronic energy, long-range quantum delocalization takes place due to resonance. However, no classical funneling occurs due to the absence of a downhill energy gradient, $\Delta(t\rightarrow 100\pi)=\sqrt{\ave{x^2}}\approx(\sum_{k=1}^{7}(k-1)^{2}\frac{1}{7})^{1/2}\approx 3.6$, as shown in figure~\ref{figure7}(c). On the other hand, when the energy gradient is comparable to, or larger in magnitude than the thermal energy $k_{B}T$, downhill classical funneling takes place, but non-resonance induces coherent localization, leading to short-range quantum delocalization, as shown in figure~\ref{figure7}(d). In the presence of a shared phonon bath (as shown in figure~\ref{figure7}(b)), however, a time-dependent Lamb shift takes place such that a near-flat energy landscape evolves into a downhill landscape via phonon relaxation. Here we take $\hbar\Omega_{j}+\hbar\lambda_{j}>\hbar\Omega_{j+1}+\hbar\lambda_{j+1}$, so that molecule $j$ has higher energy level compared to neighboring molecule $j+1$ after phonon relaxation, {\it i.e.}~$\gamma_{j}t\gg 1$ and $\gamma_{j+1}t\gg 1$, and then take $s_{j}<s_{j+1}$, so that the molecule $j$ has lower energy level compared to molecule $j+1$ at the initial time $t=0$, {\it i.e.}~$\hbar\Omega_{j}+\hbar\lambda_{j}(1+2s_{j})<\hbar\Omega_{j+1}+\hbar\lambda_{j+1}(1+2s_{j+1})$. The effective energy landscape of the linear chain then evolves from a slightly uphill landscape to a downhill landscape via a near-flat landscape, as the time-dependent Lamb shift in the effective energy landscape decays, {\it i.e.}~$\hbar\Omega_{j}+\hbar\lambda_{j}(1+2s_{j})\rightarrow\hbar\Omega_{j}+\hbar\lambda_{j}(1+2s_{j}e^{-\gamma_{j}t})\rightarrow \hbar\Omega_{j}+\hbar\lambda_{j}$ as $t\rightarrow\infty$. In this case, long-range quantum delocalization and subsequent classical funneling can be utilized for enhancing the speed of the overall exciton transfer, as shown in figure~\ref{figure7}(e): here the dotted and dashed lines display the root mean square displacement $\Delta(t)$ in figures~\ref{figure7}(c) and (d) for comparison. This suggests that sharing phonons between molecules not only enhances the coherent timescale of exciton transfer, as shown in previous studies~\cite{Nazir2009,McCutcheon2011,Wu2010,Ishizaki2010,HosseinNejad2010,Nalbach2010}, but also enables the utilizing of quantum coherence for efficient energy transport in disordered systems.

\begin{figure}
\fl\qquad\qquad\quad\includegraphics{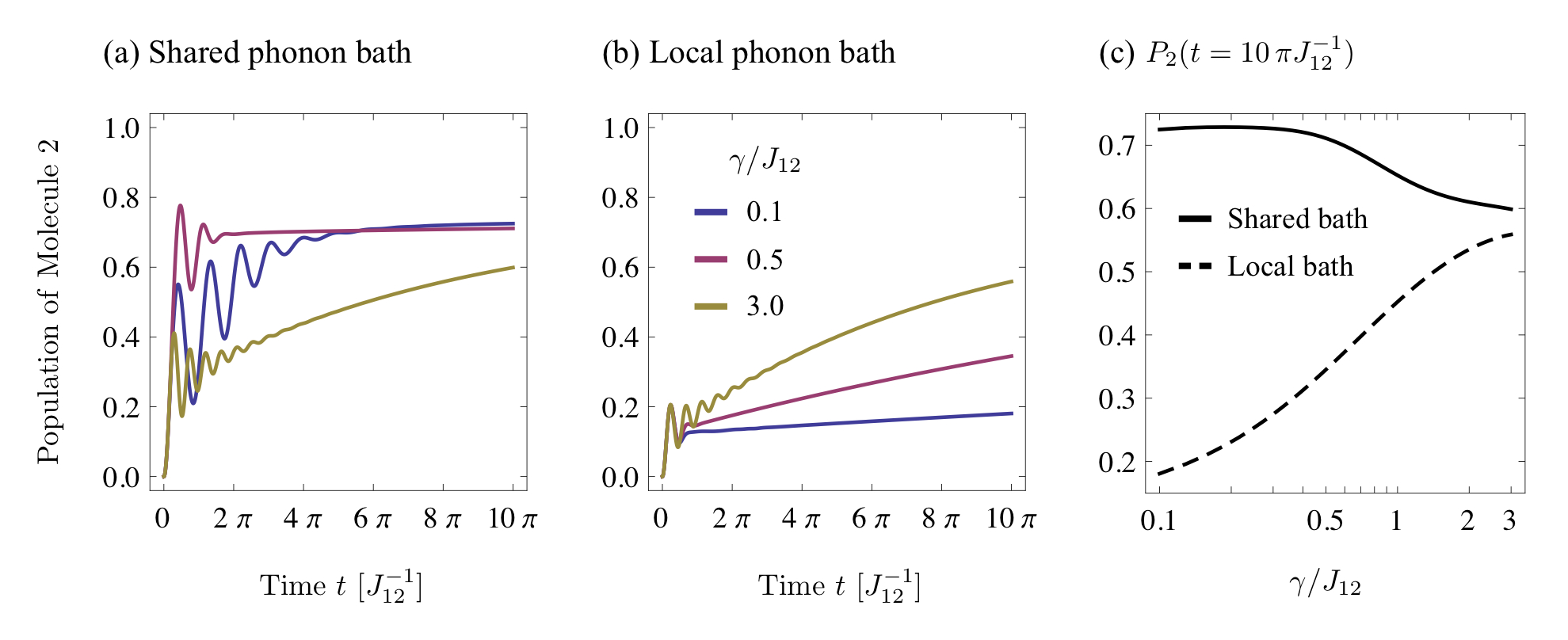}
	\caption{Dynamic resonance and phonon relaxation. Here we consider a two-molecule system and take $\hat{\rho}_{e}(0)=\ket{1}\bra{1}$, $(\Omega_1-\Omega_2)/J_{12}=4$, $k_{B}T/\hbar J_{12}=4$ and $\lambda_{j}/J_{12}=0.1$. In (a), where two molecules are coupled to a shared phonon bath with $s_{j}=\frac{3}{4}\,j\times(\Omega_1-\Omega_2)/\lambda_{j}$, the time evolution of the population of molecule 2 is shown for various phonon relaxation rates $\gamma_{j}\equiv \gamma$: here we take $\gamma/J_{12}=0.1$ (blue), 0.5 (purple) and 3.0 (brown). In (b), where the molecules are coupled to local phonon baths instead with $s_{j}=0$, the time evolution of the population of molecule 2 is shown. In (c), the population of molecule 2 at time $t=10\,\pi J_{12}^{-1}$ is displayed as a function of the phonon relaxation rate $\gamma$ for the shared phonon bath considered in (a) ({\it cf.}~solid line) and the local phonon baths considered in (b) ({\it cf.}~dashed line).}
	\label{figure8}
\end{figure}

We note that dynamic resonance enhances energy transfer in disordered systems when the phonon relaxation rate $\gamma$ is smaller than or comparable to the dipolar coupling $J$, {\it i.e.}~$\gamma\lesssim J$. In this case, the time-dependent Lamb shift induced by a shared phonon bath, $2\hbar\lambda_{j}s_{j}e^{-\gamma_{j}t}$, decays slow enough, such that an exciton resonantly hops from a donor to an acceptor while the molecules are temporally resonant. When $\gamma> J$, on the other hand, the time-dependent Lamb shift decays too quickly and disordered systems become non-resonant before the exciton hopping takes place. In figure~\ref{figure8}(a), as an example, we consider a two-molecule system coupled to a shared phonon bath where a donor has higher electronic energy than an acceptor, {\it i.e.}~$\Omega_1-\Omega_2 > J$. Here the coherent exciton transfer from the donor to the acceptor is enhanced by dynamic resonance when $\gamma\lesssim J$. The overall exciton transfer is then suppressed as the phonon relaxation rate is increased ({\it cf.}~solid line in figure~\ref{figure8}(c)). In figure~\ref{figure8}(b), on the other hand, where a two-molecule system is coupled to local phonon baths, coherent localization takes place due to the non-resonance between molecules. Here subsequent classical transport becomes faster as the phonon relaxation rate is increased and the overall exciton transfer is enhanced as the relaxation rate is increased ({\it cf.}~dashed line in figure~\ref{figure8}(c)).

\section{Asymmetric coupling structure}

Recently the present authors showed that some shared bath models considered in previous studies~\cite{Ishizaki2010,HosseinNejad2010} are equivalent to a local bath model, where each molecule is coupled to an independent phonon bath~\cite{Lim2013}. Here we call two bath models equivalent if they lead to the same time evolution of the reduced electronic state of the molecules, which describes exciton transfer dynamics. The equivalence mainly stems from the symmetric coupling structure between shared phonons and molecules~\cite{Lim2013}. Contrary to previous studies, in this work we have considered an asymmetric coupling structure where different molecules are coupled to a shared phonon mode (or a shared bath) with different coupling strengths. In section~\ref{section_nonequilibrium_regime}, where a shared phonon mode was considered, asymmetric couplings $d_1\neq d_2$ led to dynamic resonance, whereas their symmetry $d_1 = d_2$ did not lead to a time-dependent Lamb shift, as shown in equation~\eref{eq:Dynamic_L}. In section~\ref{section_quasiequilibrium_regime}, where a shared phonon bath was considered, an asymmetric coupling structure, {\it i.e.} $\lambda_{j}s_{j}e^{-\gamma_{j}t}\neq \lambda_{k}s_{k}e^{-\gamma_{k}t}$ for $j\neq k$, led to a non-trivial time-dependent Lamb shift which led to dynamic resonance. On the other hand, when $\lambda_{j}s_{j}e^{-\gamma_{j}t}=\lambda_{k}s_{k}e^{-\gamma_{k}t}$ for $j\neq k$, all the energy levels of the molecules fluctuate in the same way, where the energy-level mismatch between molecules does not depend on time $t$, leading to no dynamic resonance.

\section{Generalization to multi-level nanostructured systems}

We have considered two-level systems coupled to a shared phonon in the nonequilibrium regime. One may extend our work to nanostructures, such as optomechanical systems where optical cavity modes are coupled to a mechanical oscillator (movable mirror). Here the frequency $\omega$ of the mechanical oscillator ranges from $10\,{\rm MHz}$ to $1\,{\rm GHz}$, while the damping rate $\gamma$ of the oscillator is about $1\,{\rm kHz}$ in a cryogenic scenario~\cite{Aspelmeyer2013}. This corresponds to a Q factor of $10^{4}\sim 10^{6}$, implying that a mechanical oscillator can be in a non-equilibrium state over its natural period. When two coupled cavity modes share such a mechanical oscillator, their Hamiltonian is given by
\begin{eqnarray}
	\hat{H}&=\hbar\Omega_1\hat{a}^{\dagger}_{1}\hat{a}_{1}+\hbar\Omega_2\hat{a}^{\dagger}_{2}\hat{a}_{2}+\hbar J(\hat{a}^{\dagger}_{1}\hat{a}_{2}+\hat{a}^{\dagger}_{2}\hat{a}_{1})+\hbar\omega\hat{b}^{\dagger}\hat{b}\nonumber\\
	&\quad+\hbar(g_{1}\hat{a}^{\dagger}_{1}\hat{a}_{1}-g_{2}\hat{a}^{\dagger}_{2}\hat{a}_{2})\otimes(\hat{b}^{\dagger}+\hat{b}),
	\label{eq:optomechanics}
\end{eqnarray}
where $\Omega_{j}$ are frequencies of the cavity modes $j$ with creation and annihilation operators $\hat{a}^{\dagger}_{j}$ and $\hat{a}_{j}$, $J$ their coupling, $\omega$ a frequency of the mechanical oscillator with creation and annihilation operators $b^{\dagger}$ and $b$, and $g_{j}$ a vacuum optomechanical coupling constant. The interaction Hamiltonian between the oscillator and the cavity modes ({\it cf.}~the last term in equation~\eref{eq:optomechanics}) stems from the modulation of cavity resonance frequencies by the mechanical oscillator~\cite{Aspelmeyer2013}. Assuming the oscillator is placed in between the two cavities, one may alter the cavity-mode coupling $J$ through manipulating the oscillator's reflectance. One may also increase the coupling strength between the oscillator and the multi-level cavity modes by controlling the average number of photons in the cavities. As a result, the Hamiltonian in equation~\eref{eq:optomechanics} may be regarded as a generalization of our DRET model to multi-level systems. One could also consider polar molecules trapped in an optical lattice as a potential candidate where DRET may be applied, for which the coupling between polar molecules is comparable in magnitude to the coupling between polar molecules and shared phonon~\cite{Felipe2011}.

\section{Summary}

We investigated a shared phonon in the nonequilibrium regime and its influence on the exciton transfer in a chain of molecules. We showed that the time-dependent Lamb shift induced by phonon dynamics leads to interesting energy transport features, such as dynamic resonance, directionality and quantum walk. This work opens up some intriguing possibilities for quantum transport in disordered systems in general. Contrary to the widely accepted view that quantum coherence is not useful for energy transport in disordered systems, our results show that phonons can enable a quantum speed-up in energy transport, regardless of significant disorder. The consideration of more complex shared phonon dynamics may lead to new transport mechanisms like DRET being discovered, providing features that are unavailable in systems investigated so far. Thus, we expect our findings to help open up the engineering of novel quantum transport schemes in disordered nanostructures.

\ack
We are grateful to M S Kim, M B Plenio and S F Huelga for useful comments. This work was supported by National Research Foundation of Korea grants funded by the Korean Government (Ministry of Education, Science and Technology; grant numbers 2010-0018295 and 2010- 0015059), the UKʼs Engineering and Physical Sciences Research Council, the Leverhulme Trust, the EU STREP PAPETS and an Alexander von Humboldt Professorship.

\appendix

\section{Derivation of time-dependent Lamb shift}

Here we derive an effective Hamiltonian $\hat{H}^{\rm eff}(t)$ that gives exactly the same exciton transfer dynamics, described by the state $\hat{\rho}_{e}(t)$, as obtained by the original Hamiltonian $\hat{H}$,
\begin{eqnarray}
\fl\qquad\hat{\rho}_{e}(t)&={\rm Tr_{ph}}[\exp(-\frac{i}{\hbar}\hat{H}t)\hat{\rho}(0)\exp(\frac{i}{\hbar}\hat{H}^{\dagger}t)]\\
\fl	&={\rm Tr_{ph}}\left[{\cal T}\exp\left(-\frac{i}{\hbar}\int_{0}^{t}\hat{H}^{\rm eff}(t')dt'\right)\hat{\rho}(0){\cal T}^{\dagger}\exp\left(\frac{i}{\hbar}\int_{0}^{t}\hat{H}^{{\rm eff}\dagger}(t')dt'\right)\right],
\end{eqnarray}
where ${\cal T}$ and ${\cal T}^{\dagger}$ denote the chronological and anti-chronological time ordering operators. The effective Hamiltonian consists of three parts $\hat{H}^{\rm eff}(t)=\hat{H}^{\rm eff}_{e}(t)+\hat{H}_{\rm ph}+\hat{H}^{\rm eff}_{e-{\rm ph}}$, where $\hat{H}^{\rm eff}_{e}(t)$ is an effective Hamiltonian of the electronic states of the molecules
\begin{equation}
\fl\qquad\hat{H}^{\rm eff}_{e}(t)=\sum_{j=1}^{N}\hbar(\Omega_{j}+\lambda_{j}+\Omega_{{\rm LS},j}(t))\ket{j}\bra{j}+\sum_{j\neq k}\hbar J_{jk}\ket{j}\bra{k},
	\label{eq:H_eff_e}
\end{equation}
where $\Omega_{{\rm LS},j}(t)$ is a time-dependent Lamb shift, 
\begin{equation}
\fl\qquad\Omega_{{\rm LS},j}(t)=2s_{j}\sum_{\xi}g_{j\xi}^{2}\omega_{\xi}^{-1}\cos(\omega_{\xi}t),
	\label{eq:LS_analytic}
\end{equation}
and $\hat{H}^{\rm eff}_{e-{\rm ph}}$ is an effective Hamiltonian describing decoherence of the electronic coherence
\begin{equation}
\fl\qquad\hat{H}^{\rm eff}_{e-{\rm ph}}=\sum_{j=1}^{N}\sum_{\xi}\hbar g_{j\xi}\ket{j}\bra{j}\otimes(\hat{b}^{\dagger}_{\xi}+\hat{b}_{\xi}).
	\label{eq:non-Markov_approaches}
\end{equation}
The effective Hamiltonian provides the analytic form of the time-dependent Lamb shift induced by shared phonon dynamics. It also enables one to employ well-developed non-Markovian approaches~\cite{Makri1995JCP,IshizakiJCP2009HCME}, which have been developed to describe the interaction between molecules and phonons in the form of equation~\eref{eq:non-Markov_approaches}, as shown below. We derive the effective Hamiltonian $\hat{H}^{\rm eff}(t)$ and the time-dependent Lamb shift of $\hat{H}^{\rm eff}_{e}(t)$ with the use of a path integral representation of the reduced electronic state $\hat{\rho}_{e}(t)$ and an influence functional developed by Feynman and Vernon~\cite{Feynman1963AP}.

To implement a path integral representation, we start with the formal expression of the reduced electronic state of molecules $\hat{\rho}_{e}(t)$
\begin{equation}
\fl\qquad\hat{\rho}_{e}(t)={\rm Tr_{ph}}\left[\exp(-\frac{i}{\hbar}\hat{H}t)\hat{\rho}(0)\exp(\frac{i}{\hbar}\hat{H}^{\dagger}t)\right].
\end{equation}
By splitting the time $t$ into $M$ divisions, {\it i.e.} $t=M\Delta t$, the single-exciton state at time $t$ can be expressed as
\begin{eqnarray}
\fl\qquad\hat{\rho}_{e}(t)&={\rm Tr_{ph}}\left[e^{-\frac{i}{\hbar}\hat{H}\Delta t}\times\cdots\times e^{-\frac{i}{\hbar}\hat{H}\Delta t}\hat{\rho}(0)e^{\frac{i}{\hbar}\hat{H}^{\dagger}\Delta t}\times\cdots\times e^{\frac{i}{\hbar}\hat{H}^{\dagger}\Delta t}\right].
\end{eqnarray}
We then substitute an identity operator $\openone=\sum_{k=1}^{N}\ket{k}\bra{k}$ of the single-exciton subspace at the front and back of each time evolution operator $\exp(-\frac{i}{\hbar}\hat{H}\Delta t)$ or $\exp(\frac{i}{\hbar}\hat{H}^{\dagger}\Delta t)$ for a time increment $\Delta t$
\begin{eqnarray}
\fl\qquad\hat{\rho}_{e}(t)&=\sum_{k_{M}^{+}=1}^{N}\sum_{k_{M}^{-}=1}^{N}\ket{k_{M}^{+}}\bra{k_{M}^{-}}\sum_{k_{0}^{+}=1}^{N}\sum_{k_{0}^{-}=1}^{N}\cdots\sum_{k_{M-1}^{+}=1}^{N}\sum_{k_{M-1}^{-}=1}^{N}\nonumber\\
\fl\qquad&\quad\times{\rm Tr_{ph}}\left[\left(\prod_{l=0}^{M-1}\bra{k_{l+1}^{+}}\exp(-\frac{i}{\hbar}\hat{H}\Delta t)\ket{k_{l}^{+}}\right)\right.\nonumber\\
\fl\qquad&\quad\times\left.\bra{k_{0}^{+}}\hat{\rho}(0)\ket{k_{0}^{-}}\left(\prod_{r=0}^{M-1}\bra{k_{r}^{-}}\exp(\frac{i}{\hbar}\hat{H}^{\dagger}\Delta t)\ket{k_{r+1}^{-}}\right)\right]\label{eq:partial_trace},
\end{eqnarray}
where the superscripts $\pm$ and subscripts $l$ of the $k_{l}^{\pm}$ with non-negative integers $l\in\{0,1,\cdots,M\}$ are introduced to distinguish identity operators in different positions. The partial trace over phonon degrees of freedom ${\rm Tr_{ph}}$ in equation~\eref{eq:partial_trace} can be carried out with the use of an influence functional~\cite{Feynman1963AP}. The reduced electronic state is then expressed as
\begin{eqnarray}
\fl\qquad\hat{\rho}_{e}(t)&=\sum_{k_{M}^{+}=1}^{N}\sum_{k_{M}^{-}=1}^{N}\ket{k_{M}^{+}}\bra{k_{M}^{-}}\sum_{k_{0}^{+}=1}^{N}\sum_{k_{0}^{-}=1}^{N}\cdots\sum_{k_{M-1}^{+}=1}^{N}\sum_{k_{M-1}^{-}=1}^{N}{\cal I}(k_{0}^{+},k_{0}^{-},\cdots,k_{M}^{+},k_{M}^{-})\nonumber\\
\fl\qquad&\quad\times\left(\prod_{l=0}^{M-1}\bra{k_{l+1}^{+}}\exp(-\frac{i}{\hbar}\hat{H}_{e}\Delta t)\ket{k_{l}^{+}}\right)\nonumber\\
\fl\qquad&\quad\times\bra{k_{0}^{+}}\hat{\rho}_{e}(0)\ket{k_{0}^{-}}\left(\prod_{r=0}^{M-1}\bra{k_{r}^{-}}\exp(\frac{i}{\hbar}\hat{H}^{\dagger}_{e}\Delta t)\ket{k_{r+1}^{-}}\right),
\end{eqnarray}
where ${\cal I}(k_{0}^{+},k_{0}^{-},\cdots,k_{M}^{+},k_{M}^{-})$ is the so-called influence functional that is responsible for the influence of the phonon bath on the electronic system dynamics. In the continuous limit, {\it i.e.} $\Delta t\rightarrow 0$ or equivalently $M\rightarrow\infty$, the influence functional ${\cal I}=\lim_{\Delta t\rightarrow 0}{\cal I}(k_{0}^{+},k_{0}^{-},\cdots,k_{M}^{+},k_{M}^{-})$ can be decomposed into two parts ${\cal I}={\cal I}_{D}\times{\cal I}_{\rm LS}$ where the first term ${\cal I}_{D}$ is responsible for environmental decoherence of the electronic coherence
\begin{eqnarray}
\fl\qquad{\cal I}_{D}=&\prod_{j=1}^{N}\exp\left\{-\int_{0}^{t}dt'\left[\delta(j,k^{+}(t'))-\delta(j,k^{-}(t'))\right]\right.\nonumber\\
\fl\qquad&\qquad\quad~~\left.\times\int_{0}^{t'}dt''\left[\alpha_{j}(t'-t'')\delta(j,k^{+}(t''))-\alpha_{j}^{*}(t'-t'')\delta(j,k^{-}(t''))\right]\right\},
\end{eqnarray}
and the second term ${\cal I}_{\rm LS}$ is responsible for dynamic resonance
\begin{equation}
\fl\qquad{\cal I}_{\rm LS}=\prod_{j=1}^{N}\exp\left\{-i\int_{0}^{t}dt'\left[\delta(j,k^{+}(t'))-\delta(j,k^{-}(t'))\right]\left[\Omega_{{\rm LS},j}(t')-2s_{j}\lambda_{j}\right]\right\},
\end{equation}
leading to the time-dependent Lamb shift $\Omega_{{\rm LS},j}(t)$ in equation~\eref{eq:LS_analytic}. Note that the subscript of $k_{l}^{\pm}$ for a finite time increment $\Delta t$ is replaced with a continuous variable $t'$ or $t''$ such that $k_{l}^{\pm}\rightarrow k^{\pm}(t_l)$ with $t_l=l\Delta t$ as $\Delta t\rightarrow 0$. Here, $\delta(j,k)$ is the Kronecker delta defined by $\delta(j,k)=1$ if $j=k$ and $\delta(j,k)=0$ otherwise, and $\alpha_{j}(\tau)$ is a bath response function characterized by the spectral density $\Lambda_{j}(\omega)=\sum_{\xi}\hbar g_{j\xi}^{2}\delta(\omega-\omega_{\xi})$
\begin{eqnarray}
\fl\qquad\alpha_{j}(\tau)&=\sum_{\xi}g_{j\xi}^{2}\left[\coth(\frac{\hbar\omega_{\xi}\beta}{2})\cos(\omega_{\xi}\tau)-i\sin(\omega_{\xi}\tau)\right]\\
\fl\qquad&=\frac{1}{\hbar}\int_{0}^{\infty}d\omega\Lambda_{j}(\omega)\left[\coth(\frac{\hbar\omega\beta}{2})\cos(\omega\tau)-i\sin(\omega\tau)\right].
\end{eqnarray}
Hence, in the continuous limit $\Delta t\rightarrow 0$, the reduced electronic state can be expressed as
\begin{eqnarray}
\fl\qquad\hat{\rho}_{e}(t)&=\sum_{k_{M}^{+}=1}^{N}\sum_{k_{M}^{-}=1}^{N}\ket{k_{M}^{+}}\bra{k_{M}^{-}}\sum_{k_{0}^{+}=1}^{N}\sum_{k_{0}^{-}=1}^{N}\cdots\sum_{k_{M-1}^{+}=1}^{N}\sum_{k_{M-1}^{-}=1}^{N}{\cal I}_{D}(k_{0}^{+},k_{0}^{-},\cdots,k_{M}^{+},k_{M}^{-})\nonumber\\
\fl\qquad&\quad\times\left(\prod_{l=0}^{M-1}\bra{k_{l+1}^{+}}\exp(-\frac{i}{\hbar}\hat{H}^{\rm eff}_{e}(t_l)\Delta t)\ket{k_{l}^{+}}\right)\nonumber\\
\fl\qquad&\quad\times\bra{k_{0}^{+}}\hat{\rho}_{e}(0)\ket{k_{0}^{-}}\left(\prod_{r=0}^{M-1}\bra{k_{r}^{-}}\exp(\frac{i}{\hbar}\hat{H}^{{\rm eff}\dagger}_{e}(t_r)\Delta t)\ket{k_{r+1}^{-}}\right),
\end{eqnarray}
with $t_l=l\Delta t$ and $t_r=r\Delta t$, $\hat{H}_{e}^{\rm eff}(t)$ is the effective Hamiltonian of the electronic states of the molecules including time-dependent Lamb shift in equation~\eref{eq:H_eff_e}. This implies that the effective Hamiltonian $\hat{H}^{\rm eff}(t)$ gives exactly the same exciton transfer dynamics $\hat{\rho}_{e}(t)$ as obtained by the original total Hamiltonian $\hat{H}$. For an arbitrary form of the spectral densities $\Lambda_{j}(\omega)$, the dynamics of the reduced electronic state $\hat{\rho}_{e}(t)$ can be calculated using a path-integral approach, known as the iterative real-time quasiadiabatic propagator path-integral (QUAPI) scheme~\cite{Makri1995JCP}. In this work, however, we take an Ohmic spectral density with a Lorentz-Drude cutoff function in equation~\eref{eq:LD} to employ the so-called hierarchical approach, which is a computationally efficient framework to simulate exciton transfer dynamics~\cite{IshizakiJCP2009HCME}. The time-dependent Lamb shift then becomes $\Omega_{{\rm LS},j}(t)=2s_{j}\lambda_{j}e^{-\gamma_{j}t}$. This implies that the shared phonon bath induces dynamic resonance such that the effective energy landscape that an exciton feels during transport changes in time $\hbar\Omega_{j}+\hbar\lambda_{j}(1+2s_{j}e^{-\gamma_{j}t})$.

\section{Hierarchically coupled master equations in the presence of a shared phonon bath}

Here we provide a non-Markovian quantum master equation, represented by a collection of hierarchically coupled master equations, that reliably describes exciton transfer dynamics in the presence of a shared phonon bath.

With the shared phonon bath modeled by the Ohmic spectral density and the high temperature condition characterized by $\beta\hbar\gamma_{j}<1$ taken~\cite{IshizakiJCP2009HCME}, the time evolution of the reduced electronic state $\hat{\rho}_{e}(t)$ can be modeled by the following hierarchically coupled master equations
\begin{eqnarray}
\fl\qquad\frac{d}{dt}\hat{\sigma}(\vec{n},t)
	&=-\frac{i}{\hbar}\left[\hat{H}_{e}^{\rm eff}(t),\hat{\sigma}(\vec{n},t)\right]-\sum_{j=1}^{N}n_{j}\gamma_{j}\hat{\sigma}(\vec{n},t)\nonumber\\
	&\quad\,+\sum_{j=1}^{N}\hat{\Phi}_{j}[\hat{\sigma}(\vec{n}_{j+},t)]+\sum_{j=1}^{N}n_{j}\hat{\Theta}_{j}[\hat{\sigma}(\vec{n}_{j-},t)],
	\label{eq:hierarchical}
\end{eqnarray}
where $\hat{H}_{e}^{\rm eff}(t)$ is the effective Hamiltonian of the electronic states in equation~\eref{eq:H_eff_e} and $\vec{n}=\{n_{1},\cdots,n_{N}\}$ with non-negative integers $n_{j}\in\{0,1,\cdots\}$. The vectors $\vec{n}_{j\pm}$ differ from $\vec{n}$ only by changing $n_{j}$ to $n_{j}\pm 1$, {\it i.e.} $\vec{n}_{j\pm}=\{n_{1},\cdots,n_{j}\pm 1,\cdots,n_{N}\}$. In the above, the reduced electronic state is the lowest rank operator $\hat{\rho}_{e}(t)=\hat{\sigma}(\vec{0},t)$ with $\vec{0}=\{0,\cdots,0\}$. The other auxiliary operators $\hat{\sigma}(\vec{n},t)$ are introduced in order to take into account environmental decoherence of electronic coherence. Their dimensions are the same as that of the reduced electronic state $\hat{\rho}_{e}(t)$. At initial time $t=0$, just after an exciton is created in a molecular chain, $\hat{\sigma}(\vec{0},0)=\hat{\rho}_{e}(0)$ and all the other auxiliary operators are set to be zero, {\it i.e.} $\hat{\sigma}(\vec{n},0)=0$ for all $\vec{n}\neq \vec{0}$. The last two terms in the coupled master equations produce a coupling between the operators of different ranks $N_{\rm rank}=\sum_{j=1}^{N}n_{j}$, which are given by
\begin{eqnarray}
\fl\qquad\hat{\Phi}_{j}[\hat{\sigma}(\vec{n}_{j+},t)]&=i[\ket{j}\bra{j},\hat{\sigma}(\vec{n}_{j+},t)],\\
\fl\qquad\hat{\Theta}_{j}[\hat{\sigma}(\vec{n}_{j-},t)]&=i\frac{2\lambda_{j}k_{B}T}{\hbar}[\ket{j}\bra{j},\hat{\sigma}(\vec{n}_{j-},t)]+\lambda_{j}\gamma_{j}\{\ket{j}\bra{j},\hat{\sigma}(\vec{n}_{j-},t)\}.
\end{eqnarray}
The ranks of the auxiliary operators continue to infinity, which is impossible to treat computationally. In order to terminate the hierarchically coupled master equations, we set the higher rank operators of $N_{\rm rank}\ge N_{\rm cutoff}$ to be zero, {\it i.e.} $\hat{\sigma}(\vec{n},t)=0$ for $\vec{n}$ satisfying $N_{\rm rank}\ge N_{\rm cutoff}$. In numerical simulations, this means we increase $N_{\rm cutoff}$ until the dynamics of the reduced electronic state show convergence.

We note that the hierarchically coupled master equations in equation~\eref{eq:hierarchical} are different from those in Ref.~\cite{IshizakiJCP2009HCME}, where local phonon baths were considered. The local bath model corresponds to the case that $s_{j}=0$ for all $j$ where the effective Hamiltonian of the electronic states of molecules $\hat{H}^{\rm eff}_{e}(t)$ is reduced to the original Hamiltonian $\hat{H}_{e}$ as the time-dependent Lamb shift vanishes.

\end{document}